\documentclass[aps,prb,twocolumn,superscriptaddress,showpacs]{revtex4}
\usepackage{amsmath}
\usepackage{graphicx}
\usepackage{braket}
\usepackage{color}

\begin{document}

\title{Strong charge and spin fluctuations in La$_2$O$_3$Fe$_2$Se$_2$ }
\author{Guangxi Jin}
\affiliation{Key Laboratory of Quantum Information, University of Science and
Technology of China, Hefei, 230026, China}
\affiliation{Synergetic Innovation Center of Quantum Information and Quantum
  Physics, University of Science and Technology of China, Hefei, 230026, China}
\author{Yilin Wang}
\affiliation{Beijing National Laboratory for Condensed Matter Physics, and Institute of Physics,
Chinese Academy of Sciences, Beijing 100190, China}
\author{Xi Dai}
\affiliation{Beijing National Laboratory for Condensed Matter Physics, and Institute of Physics,
Chinese Academy of Sciences, Beijing 100190, China}
\author{Xinguo Ren}
\affiliation{Key Laboratory of Quantum Information, University of Science and
Technology of China, Hefei, 230026, China}
\affiliation{Synergetic Innovation Center of Quantum Information and Quantum
  Physics, University of Science and Technology of China, Hefei, 230026, China}
\author{Lixin He}
\affiliation{Key Laboratory of Quantum Information, University of Science and
Technology of China, Hefei, 230026, China}
\affiliation{Synergetic Innovation Center of Quantum Information and Quantum
  Physics, University of Science and Technology of China, Hefei, 230026, China}

\date{\today }

\begin{abstract}

The electronic structure and magnetic properties of the strongly correlated material
La$_2$O$_3$Fe$_2$Se$_2$ are studied by using both the density function theory plus $U$ (DFT+$U$)
method and the DFT plus Gutzwiller (DFT+G) variational method.
The ground-state magnetic structure of this material obtained with DFT+$U$
is consistent with recent experiments, but its band gap is significantly overestimated
by DFT+$U$, even with a small Hubbard $U$ value. In contrast, the DFT+G method
yields a band gap of 0.1 - 0.2 eV, in excellent agreement with experiment.
Detailed analysis shows that the electronic and magnetic properties of
of La$_2$O$_3$Fe$_2$Se$_2$ are strongly affected by
charge and spin fluctuations which are missing in the DFT+$U$ method.

\end{abstract}

\pacs{
71.20.-b, 
71.27.+a, 
75.30.-m, 
}
\maketitle

\section{INTRODUCTION}

Because of their close relationship with the Fe-based high-T$_c$
superconductor, there has been revived interest in iron oxychalcogenides,
$R$$_2$O$_3$$T$$_2$$X$$_2$ ($R$=rare earth element,
$T$=transition metal element, $X$=S or Se),
which have similar crystal structures. \cite{zhu10,free10}
One example is La$_2$O$_3$Fe$_2$Se$_2$ (LOFS), which was
first explored by Mayer et al.,\cite{mayer92} and considered to be a strongly correlated material
composed of the transition metal ion Fe$^{2+}$.
Analogous to its oxychalcogenides relatives, LOFS was determined to be a
semiconductor by experiment and claimed to be a Mott insulator.\cite{free10}

\begin{figure}[h]
\centering
\includegraphics[width=0.4\textwidth]{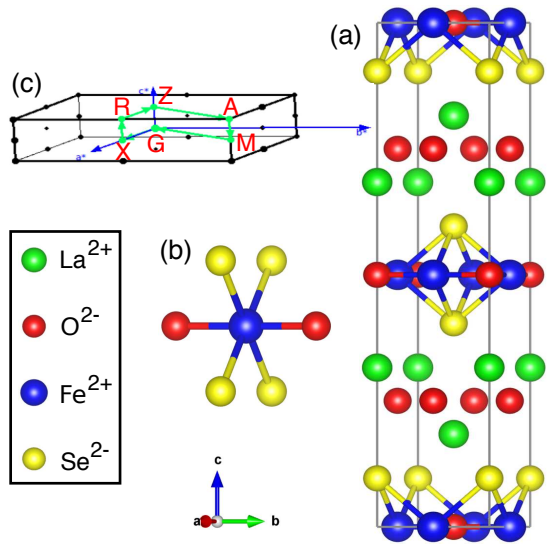}
\caption {(a). Crystal structure of LOFS, where green, red,
blue and yellow balls represent La$^{3+}$, O$^{2-}$, Fe$^{2+}$
and Se$^{2-}$, respectively;
(b) FeO$_2$Se$_4$ octahedra, where the Fe atom is surrounded
by two axial oxide ions and four equatorial selenide ions,
forming a tilted Fe-centered octahedron with the D$_{2h}$ point symmetry
(c) Symmetry points in the Brillouin zone.}
\label{fig:crystal}
\end{figure}

The crystal structure of LOFS, with space group I4/mmm
(No.139), is shown in Fig.~\ref{fig:crystal}. It is composed of alternating layered
units of [La$_2$O$_2$]$^{2+}$ and [Fe$_2$OSe$_2$]$^{2-}$, stacking along the
$c$-axis. The layered sheets of [La$_2$O$_2$]$^{2+}$, formed by
edge-sharing La$_4$O tetrahedra, expand along the $a$-$b$ plane.
The [Fe$_2$OSe$_2$]$^{2-}$ layers consist of face-sharing FeO$_2$Se$_4$
octahedra, where the Fe atom is surrounded by two axial oxygen atoms and four
equatorial selenium atoms, forming a tilted Fe-centered octahedron with the
D$_{2h}$ point symmetry. Viewed along the $c$-axis, the Fe atoms in
[Fe$_2$OSe$_2$]$^{2-}$ layer form checkerboard lattice, and the Fe-Fe
interactions are mediated by Fe-O-Fe and Fe-Se-Fe bonds.

Despite of the considerable research in the past, there are still
some mysteries about this material to be understood.
First, the magnetic structure of LOFS was found to be
anti-ferromagnetic(AFM) below the critical temperature T$_N$ $\sim$ 90 K.
However, two possible magnetic ground states have been proposed by experiments.
The first model (Model I) was proposed in Ref.~\onlinecite{free10}.
Within this model, the AFM ground state is described by the propagation vector
\textbf{k}=(0.5, 0, 0.5), and the Fe ions form a spin-frustrated
magnetic structure, which align ferromagnetially along the $a$-axis
and antiferromagnetially along the $b$-axis.
\cite{free10,landsgesell14}
An interesting aspect of this magnetic structure is that it lacks inversion symmetry, which may
further break the inversion symmetry of the crystal, resulting in ferroelectricity
by the exchange-striction effect as possible magnetic ferroelectrics.
\cite{fiebig05, cheong07} The second model (Model II)
is a non-collinear AFM model, which is composed of
two magnetic sublattices with propagation vectors $\textbf{k}_1$ = (0.5, 0, 0.5)
and $\textbf{k}_2$ = (0, 0.5, 0.5), respectively.  This magnetic structure was
first proposed by Fuwa et. al.\cite{fuwa10} for Nd$_2$O$_3$Fe$_2$Se$_2$ and was
identified as the magnetic structure for LOFS by recent experiments.  \cite{gunther14, mccabe14}
Within this model, the spins align in parallel in each sublattice, and
perpendicular between different sublattice. In contrast with Model I, the magnetic structure of Model II
still possesses the inversion symmetry.

Second, the magnitude of the local magnetic moment measured by
different experiments scatters significantly, ranging
from 2.62\ {$\mu_B$} to 3.50\ {$\mu_B$}. \cite{landsgesell14,free10,gunther14,mccabe14}
Thus information from reliable first-principles calculations will be helpful
to clarify the sitution.

Third, LOFS was determined to be a semiconductor by
electrical resistivity measurement, with a small band gap of 0.17-0.19 eV.
\cite{zhu10, lei12, landsgesell14}
However, the band gaps obtained by the DFT+$U$ method, even for very small
Hubbard $U$ parameters, are significantly larger than the experimental values.
This suggests that the Hartree-Fock type treatment of electron correlations, which
neglects the multiplet effects, might not be sufficient for this system.

In this paper, we first identify the ground-state magnetic structure of LOFS
via first-principles calculations.  We calculated the total
energies of different magnetic structures using density-functional theory
(DFT), with on-site Coulomb interaction correction (DFT+$U$),
including the two experimentally proposed magnetic structure models.
Our results suggest that Model II is the ground state-magnetic structure
of LOFS. However, the DFT+$U$ method greatly overestimates
the band gap of LOFS.~\cite{zhao13}
In order to correctly describe the electronic structure of
LOFS,\cite{zhu10, lei12} we further calculated
the electronic and magnetic properties of this compound
by the DFT plus Gutzwiller (DFT+G) method.
With appropriate $U$, $J$, parameters, the obtained DFT+G band gap is approximately 0.1 eV - 0.2 eV,
in excellent agreement with experiment. The local magnetic moment obtained by DFT+G
is about 3.0~${\mu}_B$, falling within the range of experimental results, but is somewhat smaller
than  the DFT+$U$ values, which are approximately 3.4 - 3.6~${\mu}_B$.
Detailed analysis shows that there are strong charge and spin fluctuations
in this system, which are responsible for a significant reduction of the band gap.

The rest of the paper is organized as follows. In Sec.\ref{sec:method},
the methods used in our calculations are described.
In Sec. \ref{sec:result}~A, we determine the ground-state magnetic structure of
LOFS by comparing the total energies of various spin
configurations using DFT+$U$ method. In Sec. \ref{sec:result}~B,
we study its band structure using DFT+G method. A summary of our work is given in Sec: \ref{sec:summary}.

\section{Computational details}
%
\label{sec:method}
\subsection{DFT+$U$}
We perform first-principles calculations based on
DFT within the spin-polarized generalized gradient approximation
(SGGA) using  Perdew-Burke-Ernzerhof functional,
\cite{perdew96} implemented in the Vienna ab initio 
simulations package (VASP).\cite{kresse93,kresse96}
The projector-augmented-wave (PAW) pseudopotentials with a 500 eV plane-wave
cutoff are used. To account for the correlation effect of Fe ions, we add on-site
Coulomb interaction $U$ as is done in the DFT+$U$ scheme.\cite{anisimov91}
The total energies are converged to 10$^{-8}$ eV.

\subsection{DFT+G}

LOFS is a strongly correlated system, whose band structure understandably cannot be well
described by single-particle mean-field approximations like DFT+$U$.
As mentioned above and detailed below, the band gap obtained
from DFT+$U$ calculations are too big. To correct this, we
resort to the DFT+G variational method,\cite{bunemann98, deng09}
which can treat the multiplet effects more accurately.
The DFT+G method starts with the following many-body Hamiltonian,
\begin{equation}
\begin{split}
{\hat{H}} = & {\hat{H}}_{TB} + {\hat{H}}_{int} + {\hat{H}}_{dc} \\
= & {\hat{H}}_{TB}
   + \sum_i {\hat{H}}_{i,atom} - \sum_i {\hat{H}}_{i,dc} \, .
\end{split}
\label{eq:gw_hamilton1}
\end{equation}
The first term in Eq.~(\ref{eq:gw_hamilton1}) is a $d$-$p$ tight-binding (TB) Hamiltonian constructed from
non-spin-polarized GGA band structure, projected to the $d$-$p$ manifold of
 maximally localized Wannier functions.
\cite{marzari97,souza01,marzari12,mostofi08}
This term apparently describes the hopping of electrons.
The Wannier functions contains not only the localized 3$d$ orbitals of Fe atoms,
but also extended 2$p$ orbitals of O atoms and 4$p$ orbitals of Se atoms. Using the Wannier functions,
the TB term can be written more explicitly as,
\begin{equation}
\begin{aligned}
{\hat{H}}_{TB} = \sum_{\substack{i,j \\ {m_1},{m_2} \\ {\sigma}}}
 t_{i,j}^{{m_1}{\sigma},{m_2}{\sigma}}
 {\hat{d}}_{i{m_1}{\sigma}}^{\dag}{\hat{d}}_{j{m_2}{\sigma}}
+ \sum_{\substack{i,j \\ {m_1},{m_2} \\ {\sigma}}}
 t_{i,j}^{{m_1}{\sigma},{m_2}{\sigma}}
 {\hat{p}}_{i{m_1}{\sigma}}^{\dag}{\hat{p}}_{j{m_2}{\sigma}} \\
+ \sum_{\substack{i,j \\ {m_1},{m_2} \\ {\sigma}}}
 t_{i,j}^{{m_1}{\sigma},{m_2}{\sigma}}
 {\hat{d}}_{i{m_1}{\sigma}}^{\dag}{\hat{p}}_{j{m_2}{\sigma}}
+ \sum_{\substack{i,j \\ {m_1},{m_2} \\ {\sigma}}}
 t_{i,j}^{{m_1}{\sigma},{m_2}{\sigma}}
 {\hat{p}}_{i{m_1}{\sigma}}^{\dag}{\hat{d}}_{j{m_2}{\sigma}}
\end{aligned}
\label{eq:gw_hamilton2}
\end{equation}
%
where the operator ${\hat{d}}_{im{\sigma}}^{\dagger}$(${\hat{d}}_{im{\sigma}}$)
creates (annihilates) a 3$d$ electron of Fe atom on site $i$,
with orbital $m$ and spin ${\sigma}$.
Likewise, ${\hat{p}}_{im{\sigma}}^{\dagger}$(${\hat{p}}_{im{\sigma}}$)
creates (annihilates) a $p$ electron of O and Se atoms.

The second term in Eq.~(\ref{eq:gw_hamilton1}) is a rotationally invariant
Coulomb interaction Hamiltonian describing the strong on-site electron-electron
interactions within the 3$d$ orbitals of Fe atoms.
We assume the spherical symmetry of local environment of the Fe atom
and use a full interaction tensor as
$U_{{m_1}{\sigma},{m_2}{\sigma}^{\prime},{m_3}{\sigma}^{\prime},{m_4}{\sigma}}$.
For the detailed definition of the $U$ tensor, we follow the method described in Ref.~\onlinecite{wang15}.
Within the complex spherical harmonics basis, the second term of the Hamiltonian in Eq.~(\ref{eq:gw_hamilton1})
can be explicitly expressed as,\cite{georges13}
\begin{equation}
\begin{split}
&{\hat{H}}_{i,atom} \\
= &\sum_{\substack{{m_1},{m_2}, {m_3},{m_4} \\ {\sigma},{\sigma}^{\prime}}}
U_{{m_1}{\sigma},{m_2}{\sigma}^{\prime},{m_3}{\sigma}^{\prime},{m_4}{\sigma}}
{\hat{d}}_{{m_1}{\sigma}}^{\dag} {\hat{d}}_{{m_2}{\sigma}^{\prime}}^{\dag}
{\hat{d}}_{{m_3}{\sigma}^{\prime}} {\hat{d}}_{{m_4}{\sigma}}  \,
\end{split}
\label{eq:gw_hamilton3}
\end{equation}
where the $U$ tensor satisfies the condition,
\begin{equation}
U_{{m_1}{\sigma},{m_2}{\sigma}^{\prime},{m_3}{\sigma}^{\prime},{m_4}{\sigma}}
= {\delta}_{{m_1}+{m_2},{m_3}+{m_4}} \sum_{k}c_k^{{m_1},{m_4}}c_k^{{m_2},{m_3}}F^k  \, .
\label{eq:gw_hamilton4}
\end{equation}
Here, $m_1$, $m_2$, $m_3$ and $m_4$ are the orbital index, ${\sigma}$,
${\sigma}^{\prime}$ denote the spin states,
$c_k^{m_1,m_4}$ are the Gaunt coefficients,
and $F^k$ is the Slater integrals.
For the $d$ shell, $k$=0, 2, 4, and
hence the full $U$ tensor can be specified by the parameters $F^0$, $F^2$ and $F^4$.
According to Wang et. al.,\cite{wang15} $F^4$/$F^2$ = 0.625 is an approximation with good accuracy for
the d shell,\cite{groot90} and hence is also adopted in this work.
The intra-orbital Coulomb interaction and Hund's rule coupling
are set to be $U = F^0 + \dfrac{4}{49}F^2 + \dfrac{4}{49}F^4$,
and $J = \dfrac{5}{98}(F^2 + F^4)$, respectively.
Therefore, given the parameter values of either $F^0$, $F^2$ or $U$, $J$,
we can construct the full interaction U tensor.

The last term of Eq.~(\ref{eq:gw_hamilton1}) is a double-counting (DC) term in order
to substrate the correlation effect which has been partially included in DFT calculations.
The DC term is not uniquely defined, and here
we adopted the choice used in Ref.~\onlinecite{liechtenstein95}, where it can be expressed as
\begin{equation}
\begin{split}
{\hat{H}}_{dc} &= \sum_{\sigma} {U_{dc}^{\sigma}} {\hat{n}}_d^{\sigma} \\
U_{dc}^{\sigma} &= U(n_d-{\frac{1}{2}}) - J(n_d^{\sigma}-1)/2  \\
n_d^{\sigma} &= \sum_{m} {\langle} {\Psi}_G |
{\hat{d}}_{m{\sigma}}^{\dag} {\hat{d}}_{m{\sigma}} | {\Psi}_G {\rangle} \, .
\end{split}
\label{eq:gw_hamilton7}
\end{equation}

The Gutzwiller trial wave function $| {\Psi}_G {\rangle}$ is
constructed by applying a projection operator  ${\hat{P}}$
on the uncorrelated wave function $| {\Psi}_0 {\rangle}$ from DFT calculations,
\begin{equation}
| {\Psi}_G {\rangle} = {\hat{P}} | {\Psi}_0 {\rangle}  \,
\label{eq:gw_wf1}
\end{equation}
with
\begin{equation}
{\hat{P}} = \prod_{\textbf{R}} {\hat{P}}_{\textbf{R}} = \prod_{\textbf{R}}
\sum_{{\Gamma},{\Gamma}^{\prime}}
    {\lambda}(\textbf{R})_{{\Gamma}{\Gamma}^{\prime}} | {\Gamma},{\textbf{R}}
    {\rangle} {\langle} {\Gamma}^{\prime},{\textbf{R}} |  \,
\label{eq:gw_wf2}
\end{equation}
where $| {\Gamma},{\textbf{R}} {\rangle}$ are the eigenstates of 
the on-site Hamiltonian ${\hat{H}}_{i,atom}$ for site {\textbf{R}}, 
and  ${\lambda}(\textbf{R})_{{\Gamma}{\Gamma}^{\prime}}$ are the
Gutzwiller variational parameters to be determined by minimizing
the total-energy of the ground state ${\mid} {\Psi}_G {\rangle}$, through
the variational method.\cite{bunemann98, deng09}
More details of this method can be found in Refs.~\onlinecite{deng09} and \onlinecite{lanata12}.

\section{results and discussion}
%
\label{sec:result}

In this section, we first study the magnetic ground state of LOFS
using the DFT+$U$ method. The most stable magnetic configuration coming
out from our DFT+$U$ calculations agrees with
the one proposed by Fuwa and coworkers.\cite{fuwa10}
However, as mentioned above, the DFT+$U$ method significantly overestimates the band gap of LOFS.
We then study the band structure of LOFS using the DFT+G method.

\begin{figure}%
\centering
\includegraphics[width=0.4\textwidth]{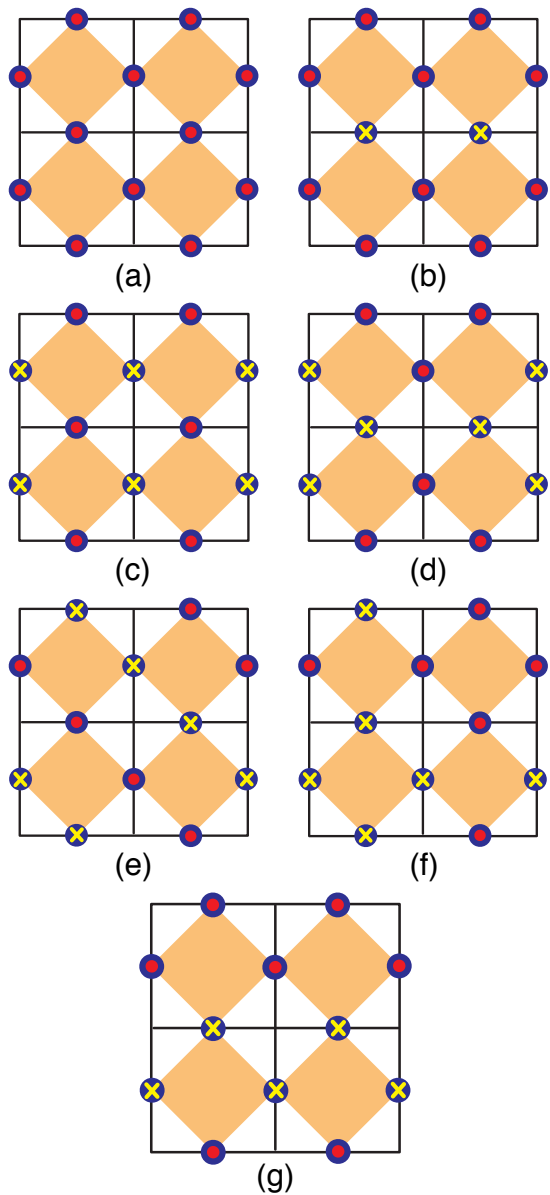}
\caption{(a)-(g) The magnetic structures of one layer of
         Fe atoms in the crystalline $ab$ plane.
         Only the Fe atoms are shown in figures, which is
         indicated by large blue circles; and the small red circle and yellow cross
         represent up and down spin orientation, respectively.}
\label{fig:magnet_stru} 
\end{figure}

\subsection{Results from DFT+$U$}

In order to identify the magnetic structure of the ground state, we performed
a series of DFT+$U$ total energy calculations for different magnetic configurations.
In LOFS, each unit cell contains two [Fe$_2$OSe$_2$]$^{2-}$
layers along the $c$-axis.
Atomic positions of the second [Fe$_2$OSe$_2$]$^{2-}$
layer are shifted by (0.25, 0.25, 0.25) of the lattice vectors relative to the
first layer.
Possible magnetic configurations in each layer are shown in Fig.~\ref{fig:magnet_stru}.
The magnetic structures in a unit cell are then the combinations of the magnetic
configurations of the two layers.
For example the configuration [(b)+(a)] means
the first layer takes configuration (b),
whereas the second layer takes configuration (a).
This notational system is similar to that used
by Zhu et. al.\cite{zhu10} and Zhao et. al.\cite{zhao13}

We calculated the total energies of seven magnetic structures,
including (1) FM[(a)+(a)], (2) AFM1[(c)+(c)], (3) AFM2[(b)+(a)],
(4) AFM3[(g)+(g)], (5) AFM4[(d)+(f)],
(6) AFM5[(e)+(e)], (7) AFM6[(d)+(d)].
In the AFM3 configuration,
the spins in each [Fe$_2$OSe$_2$]$^{2-}$ layer form a double stripe AFM structure
along the $a$-axis,~\cite{free10}, which is different from the structure used in
Ref.~\onlinecite{zhu10} and Ref.~\onlinecite{zhao13}.
AFM3 is actually the experimental magnetic structure Model I
proposed by Free et. al.,\cite{free10}
whereas the AFM6 configuration corresponds to the experimental
magnetic model II, within the collinear approximation.
Such an approximation was previously adopted for Sr$_2$F$_2$Fe$_2$OS$_2$
in Ref.~\onlinecite{zhao13}, because the magnetic-anisotropy energies are
rather small compared to the energy differences between different configurations
(see below).

Experimentally it was found that the spins form AFM along the $c$-axis.
We calculate the total energy of spin configurations with propagation vectors
(1/2,0,1/2) and (1/2,0,0) for the experimental magnetic model I (AFM3, using a
2$\times$1$\times$2 supercell. We found that the energy difference between the AFM spin
configuration along the $c$-axis with magnetic propagation vector
${\bf k}$=(0.5, 0, 0.5), and the ferromagnetic configuration along $c$-axis
with propagation vector ${\bf k}$=(0.5, 0, 0), is only about 0.1 meV
in a 2$\times$1$\times$2 sueprcell.
Therefore, in the following studies, we ignore the anti-ferromagnetic
configuration between the unit cells along the $c$-axis, and focus on the
magnetic structure in the $ab$ plane. To accommodate
all seven magnetic structures,
we use a 2$\times$2$\times$1 supercell.
The corresponding Monkhorst $k$-mesh is set to
8$\times$8$\times$4.

The calculated energies for different magnetic configurations, are listed in
Table~\ref{tab:En_vs_mag_gga} for various effective Coulomb $U_{eff}$=$U$-$J$,
where $U$, and $J$ are the Coulomb and Hund's exchange interactions respectively .
One can see that, for all $U_{eff}$, the total energy of AFM6 is significantly
lower than those of AFM3 and other spin configurations.
The magnetic ground state of LOFS
is then determined to be AFM6 for all the values of $U_{eff}$ 
considered in our DFT+$U$ calculations.
We therefore conclude that the experimental magnetic structure Model II should be
the ground state magnetic structure in LOFS.
These results are consistent with previous DFT+$U$ calculations for Sr$_2$F$_2$Fe$_2$OS$_2$.
\cite{zhao13} We note however, in Ref.~\onlinecite{zhu10},
the ground state of LOFS was determined to be AFM6 for
$U_{eff}$=0, 1.5 and 3.0 eV, but changed to AFM1 at $U_{eff}$=4.5 eV.

\begin{table}
\caption{Relative energy $\Delta$E (meV/unit cell) of different magnetic
configurations and various parameter $U_{eff}$ (unit in eV),
with the reference energy of FM, where the crystal structure was constrained
at I4/mmm space group symmetry.}
\begin{tabular}{lccccccccc}
\hline \hline
$U_{eff}$  & FM    & AFM1      & AFM2     & AFM3      & AFM4      & AFM5     & AFM6     \\
\hline
0        & 0     &  159.65   &  15.67   & -120.09   & -40.61    & -10.04   & -128.78  \\

1.5      & 0     & -237.42   & -52.06   & -237.12   & -236.51   & -341.42  & -394.25  \\

3.0      & 0     & -220.76   & -49.52   & -194.86   & -194.88   & -273.04  & -303.11 \\

4.5      & 0     & -184.12   & -42.42   & -148.38   & -148.49   & -201.93  & -219.38 \\
\hline   
\hline
\label{tab:En_vs_mag_gga}
\end{tabular}
\end{table}

For comparison, we also calculated the total energies of different magnetic configurations
for Pr$_2$O$_3$Fe$_2$Se$_2$, which has the similar crystal structure
as LOFS.\cite{ni11}
We found that its magnetic ground state is also AFM6 within the DFT+U approximation,
the same as that of LOFS and Sr$_2$F$_2$Fe$_2$OS$_2$.
\cite{zhao13}
These results suggest that the Model II magnetic structure
should be the common character for the oxychalcogenide materials
$R_2$O$_3$Fe$_2$Se$_2$ ($R$=rare earth).

\begin{table}
\caption{Magnetic moment (magmom) of LOFS,
calculated by DFT+$U$ method, under AFM6 magnetic configuration with different $U_{eff}$.}
\begin{tabular}{ccccccc}
\hline
\hline
$U_{eff}$ (eV)   &   0    &  1.5     &  3.0     &  4.5    \\

magmom($\mu_B$)  &   -    &  3.4   &  3.5   &  3.6  \\
\hline   
\label{tab:magmom_sgga_u}
\end{tabular}
\end{table}

After determining the ground-state magnetic structure, we
calculate the magnetic moments of LOFS in the
AFM6 configuration for different values of $U_{eff}$.
The results are listed in Table~\ref{tab:magmom_sgga_u}.
The calculated magnetic moments of the Fe ion for different $U_{eff}$ values are
around 3.4~{$\mu_B$} - 3.6~{$\mu_B$}, which are in agreement with the
experimental result 3.50~{$\mu_B$}, obtained by
McCabe et. al.\cite{mccabe14}

\begin{figure}[h]
\centering
\includegraphics[width=0.4\textwidth]{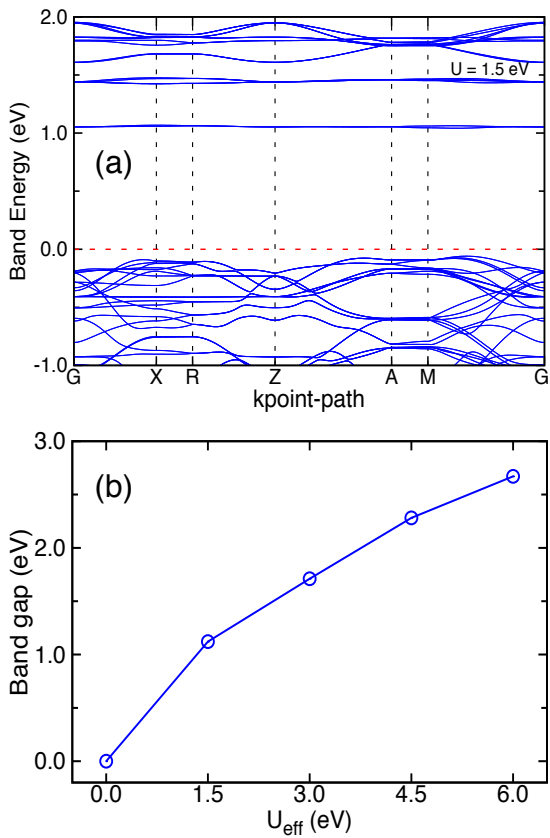}
\caption {(a)
Band structure of LOFS calculated by DFT+$U$ method,
under the ground state magnetic structure AFM6
and $U_{eff}$=1.5 eV.
 The Fermi energy have been set to 0 eV.
(b) The band gap of LOFS as a function of $U_{eff}$
calculated by DFT+$U$
method, under ground state magnetic structure.}
\label{fig:sggau_gap_u_relate}
\end{figure}

To study the electronic structure of LOFS, we calculated
the band structure and density of states (DOS) by using the
DFT+U method for the AFM6 spin configuration, using different
Coulomb $U_{eff}$=0 -- 6.0~eV.
The typical band structures of LOFS with $U_{eff}$=1.5~eV are
shown in Fig.~\ref{fig:sggau_gap_u_relate}(a).
Even for a small Coulomb $U_{eff}$=1.5~eV, the band gap is as large
as 1.12~eV, which is significantly larger than the
energy gap $E_g \sim$  0.17~eV - 0.19~eV, extracted from
the electrical resistivity measurement.\cite{zhu10,lei12}
The calculated band gaps
as a function of $U$ are shown in Fig.~\ref{fig:sggau_gap_u_relate}(b).
For $U_{eff}$=0, the system is metallic.
For $U_{eff}>$0, there is a nearly linear dependence of the band gap
upon the $U_{eff}$ value as can be seen from Fig.~\ref{fig:sggau_gap_u_relate}(b).
For a reasonable $U_{eff}$=4.5~eV, the
calculated band gap is approximately 2.0~eV,
which is about one order of magnitude larger than the experimental value.
The results suggest that the correlation effects are not accounted for adequately
by the DFT+$U$ method, and more advanced methods are needed to describe the electronic
structure of LOFS.

\begin{figure}[h]
\centering
\includegraphics[width=0.4\textwidth]{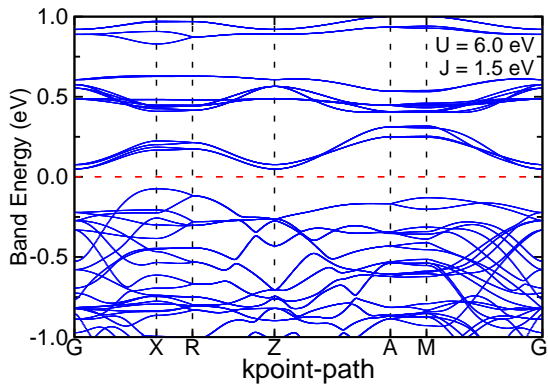}
\caption {Band structure of LOFS, calculated by
  DFT+G, with $U$=6.0 eV and $J$=0.25$U$, under ground state magnetic
  structure. The Fermi energy have been
  set to 0 eV.}
\label{fig:band_gw_U6.0J0.25}
\end{figure}
%

\subsection{Results from DFT+G}

To correctly describe the electronic structure of LOFS,
we then performed DFT+G calculations under the ground state magnetic structure AFM6.
We do calculations with a series of Hubbard $U$=3.0 - 7.0~eV and Hund's
exchange $J$=0.1~$U$ - 0.3~$U$.
We show the results here for typical parameters $U$=6.0~eV, $J$=0.25~$U$,
which corresponds to $U_{eff}$=$U$-$J$=4.5~eV in the DFT+$U$ calculations.

The band-structure calculated with typical $U$=6.0~eV and $J$=0.25~$U$
is shown in Fig.~\ref{fig:band_gw_U6.0J0.25}.
The DFT+G calculated band gap is approximately 0.121 eV,
which is in excellent agreement with the value obtained by the
electric resistivity measurement.\cite{zhu10, lei12}
This is in stark contrast with those obtained from DFT+$U$ calculations.
For example, DFT+U gives a very large band gap (approximately 2.28~eV) with $U_{eff}$=4.5~eV.
These results clearly demonstrate that the multiplets effects, which are missing
in the DFT+$U$ methods but captured in the DFT+G method, are crucial for a  correct description
of the electronic structure of LOFS.

We also calculate the magnetic moments of Fe atoms
using the Gutzwiller wave functions.
The magnetic moments of Fe atoms
are approximately 3.08~$\mu_B$ for $U$=6.0~eV and $J$=0.25~$U$.
This value is between the experimental results of
Ref.~\cite{free10} and Ref.~\cite{gunther14}, which are somehow
smaller than those obtained from DFT+$U$ calculations.

\begin{figure}[h]
\centering
\includegraphics[width=0.4\textwidth]{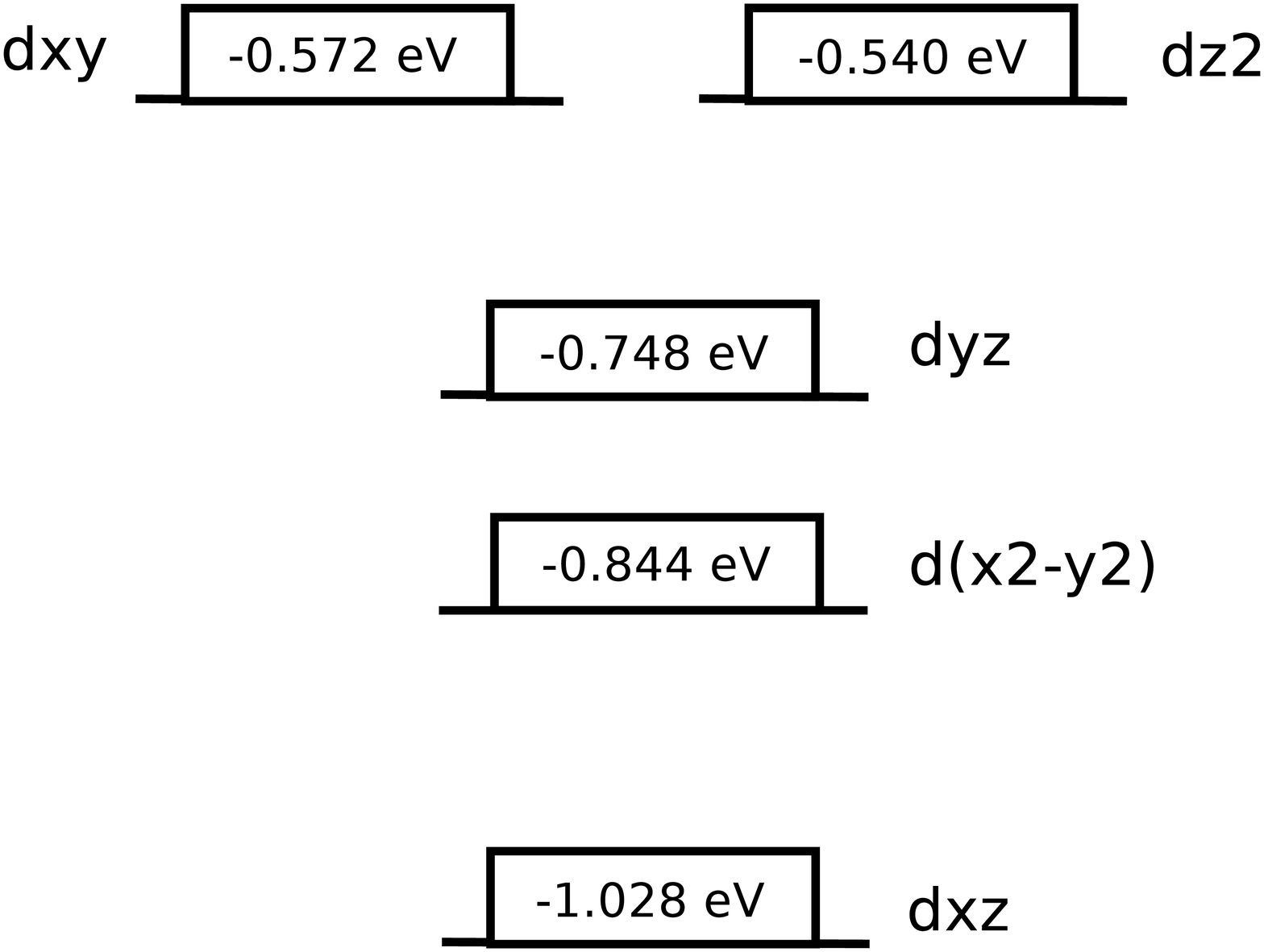}
\caption {Energy levels of Fe 3d states under crystal field splitting. }
\label{fig:crystal_field_Fe}
\end{figure}

\begin{figure}[h]
\centering
\includegraphics[width=0.4\textwidth]{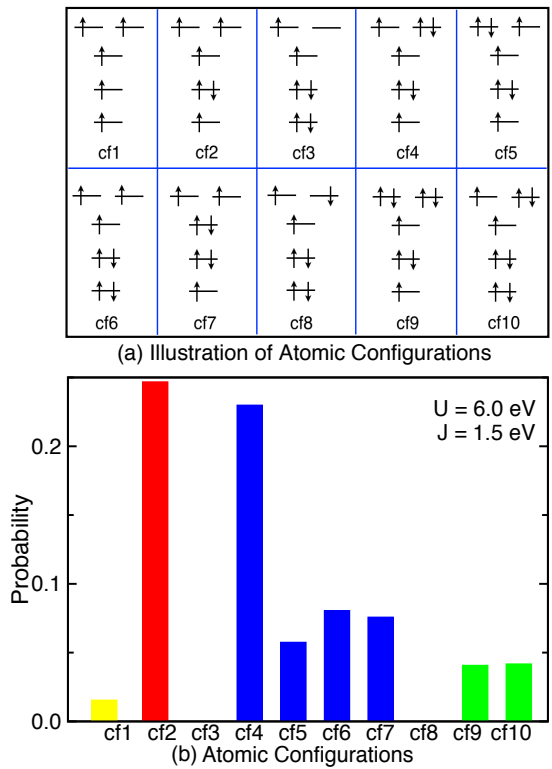}
\caption {(a) Illustration of main atomic configurations with relatively
large probability; (b) Probability of the atomic configurations $| I {\rangle}$
in the Gutzwiller wave function $| G {\rangle}$, calculated
with $U$=6.0 eV and $J$=0.25 $U$, under ground state magnetic structure.
Here, the atomic configurations with occupation number 5, 6, 7 and 8,
are represented by histograms with color yellow, red, blue and green, respectively. }
\label{fig:acf_U6.0J0.25}
\end{figure}

The differences between the DFT+$U$ and DFT+G methods are that
the DFT+G methods correctly take account of
the multiplets effects whereas in DFT+$U$ methods only a single atomic configuration is considered.
To understand the results, we further analyzed the  Gutzwiller wave
functions. We calculated the probability of the atomic multiplets
$| I \rangle$ of Fe atoms
using the Gutzwiller ground state wave function $|G \rangle$, using 
the relation $P_I$ = ${\langle} G | I {\rangle}{\langle} I | G {\rangle}$.
To display the atomic configuration more explicitly, the crystal
field splitting of Fe atom is
shown in Fig.~\ref{fig:crystal_field_Fe}.
The ten major atomic configurations with relatively large
population are shown in Fig.~\ref{fig:acf_U6.0J0.25}(a),
and corresponding populations are shown in Fig.~\ref{fig:acf_U6.0J0.25}(b)
for $U$=6.0~eV, and $J$=0.25~$U$.

First we look at the electron occupation number of these atomic configurations.
The atomic configurations cf1 has occupation number $n$=5 (yellow),
and has the population P($n$=5) = 0.015.
Configurations cf2, cf3 have occupation number 6 (red), and
their total population P($n$=6) = 0.247.
Configurations cf4, cf5, cf6, cf7, cf8 have
occupation number 7 (blue) and total population P($n$=7) = 0.4438,
and configurations cf9, cf10 have occupation number 8 (green) with
total population P($n$=8) = 0.0826.
These results suggest that there are strong charge fluctuations on
the Fe ions.
Although the local interaction in this material is quite strong leading to Mott insulator behavior,
the charge and spin fluctuation is still strong for such a multi-orbital system,
which reduces the single particle gap from the value obtained by Hatree-Fock-type approximation (i.e. DFT+$U$) to about 0.1-0.2 eV.

Besides the charge fluctuation, there are also strong spin fluctuation on the Fe atoms.
Configurations cf3, cf8, cf9, cf10 have total spin $S$=2.
The total population of these configurations is P($S$=2) = 0.0827.
Configurations cf4, cf5, cf6, cf7 have total spin $S$=3, and the their
total population is P($S$=3) = 0.4438.
Configuration cf2 has $S$=4, and P($S$=4) = 0.247,
and cf1 has $S$=5 with P($S$=5) = 0.015.
The most populated spin states in DFT+G calculations are $S$=3, which is smaller than the
formal magnetic state $S$=4 in DFT+$U$ calculations.
As a result, the DFT+G calculated magnetic moments of Fe ions are smaller
than those calculated by DFT+$U$ methods.

\section{Summary}
\label{sec:summary}

We have studied the electronic structure and magnetic properties
of the strongly-correlated material La$_2$O$_3$Fe$_2$Se$_2$,
using both DFT+$U$ and DFT+G methods.
The ground states magnetic configuration obtained from DFT+$U$ calculations
are in agreement with most recent experiments.\cite{mccabe14,gunther14}
However, DFT+$U$ calculations
greatly overestimate the band gap of the material.
We then investigate electronic structure using the DFT+G method, and the results
show La$_2$O$_3$Fe$_2$Se$_2$ is a narrow gap semiconductor, in excellent agreement
with experiments.
We show there are strong charge and spin fluctuations
on the Fe atoms that greatly reduce the band gap and magnetic moments from
the DFT+$U$ values.

\acknowledgments

LH acknowledges the support from
Chinese National Science Foundation Grant number 11374275.


\end{document}